\documentclass[aps,prb,twocolumn,showpacs, superscriptaddress]{revtex4-1}

\usepackage{graphicx}
\usepackage{dcolumn}
\usepackage{bm}
\usepackage{amsmath}

\begin{document}


\title{Controlling the nuclear polarization in quantum dots using optical pulses with a modest bandwidth}

\author{S. G. Carter}
\affiliation{Naval Research Laboratory, Washington, DC 20375-5322, USA}
\author{Sophia E. Economou}
\affiliation{Naval Research Laboratory, Washington, DC 20375-5322, USA}
\author{A. Shabaev}
\affiliation{George Mason University, Fairfax, VA 22030, USA}
\author{A. S. Bracker}
\affiliation{Naval Research Laboratory, Washington, DC 20375-5322, USA}

\date{\today}

\begin{abstract}
We show that detuned optical pulse trains with a modest spectral width can polarize nuclear spins in InAs quantum dots. The pulse bandwidth is large enough to excite a coherent superposition of both electron spin eigenstates in these negatively charged dots but narrow enough to give partial spectral selectivity between the eigenstates. The coherent precession of electron spin states and periodic excitation focuses the nuclear spin distribution, producing a discrete set of precession modes. The spectral selectivity generates a net nuclear polarization, through a mechanism that relies on optical spin rotations rather than electron spin relaxation.  
\end{abstract}

\pacs{78.67.Hc, 78.47.D-, 72.25.Fe }

 

\maketitle

\section{Introduction}

Optical pulse control is extremely versatile in controlling quantum processes in matter. The center wavelength of the pulse can be used to select particular transitions or absorbing species, and the shape and detuning of the pulse can tailor the dynamics. In semiconductor quantum dots, optical pulse control has recently enabled a great deal of progress toward using spins as quantum bits.\cite{Greilich_Science06, Carter_PRB07, Berezovsky_Science08, Press_Nature08, Carter_PRL09, Greilich_Nphys09, Kim_PRL10} Optical pulses have been used to perform single qubit gates, \textit{i.e.} spin rotations, in single QDs \cite{Berezovsky_Science08, Press_Nature08, Greilich_Nphys09, Kim_PRL10} and even to manipulate entangled states in a two spin qubit system of coupled QDs.\cite{DKim_NPhys10}

Optical pulses have also been used to address one of the major challenges in using spins in QDs as a qubit:\cite{Greilich_Science07, Carter_PRL09, Greilich_PRB09, Ladd_PRL2010} the hyperfine interaction with a large ($10^4$-$10^5$) ensemble of nuclear spins (n-spins) in each dot. The net n-spin polarization is typically random and changes over time, leading to fluctuations in the electron spin (e-spin) splitting through the Overhauser field. For e-spins the hyperfine interaction is particularly strong, leading to inhomogeneous dephasing times $T_2^*$ of a few nanoseconds.\cite{Merkulov_PRB02} Narrowing the distribution of n-spin polarizations can increase $T_2^*$ and perhaps even the true decoherence time $T_2$,\cite{Stepanenko_PRL06, Greilich_Science07, Reilly_Science08, Xu_Nature09, Vink_NPhys09, Latta_NPhys09, Greilich_PRB09} and completely polarizing the nuclei has been predicted to increase $T_2$,\cite{taylor} which is also limited by the hyperfine interaction.\cite{Khaetskii_PRL02, Coish_PSS09} Many experiments have now demonstrated large nuclear polarizations through optical pumping of e-spins,\cite{Brown_PRB96, Bracker_PRL05, Braun_PRB06, Makhonin_APL08, Krebs_PRL10} but narrowing the distribution of n-spin polarizations and observing an improvement of e-spin coherence has been challenging.


In Ref.~\onlinecite{Greilich_Science07}, a train of optical pulses resonant with an ensemble of QDs was observed to stabilize the nuclear polarization at a discrete set of values related to the pulse repetition frequency. Electron spins precessing in an external magnetic field, which was perpendicular to the optical axis, were repeatedly excited by the pulse train unless their precession frequency was synchronized to a multiple of the repetition rate. Repeated excitation of unsynchronized spins led to rapid n-spin flips with no preferred direction, producing a random walk in the nuclear polarization and precession frequency until finding a synchronized frequency. In Ref.~\onlinecite{Carter_PRL09}, detuned pulses were used to generate a component of the spin vector along the external magnetic field, producing directional n-spin flips and more stable nuclear polarization distribution. However, in each of these experiments the net nuclear polarization was expected to be near zero since only a small change in the nuclear polarization will induce synchronization.

Another approach toward optically narrowing the n-spin distribution has focused on resonant continuous-wave (cw) excitation of transitions from the spin eigenstates with high spectral selectivity.\cite{Korenev_PRL07, Latta_NPhys09, Xu_Nature09} Recent experiments have shown locking of the optical transition to the laser through nuclear feedback that polarizes and narrows the n-spin distribution.\cite{Latta_NPhys09, Xu_Nature09} In Ref.~\onlinecite{Latta_NPhys09}, the laser quickly polarizes the e-spins by pumping out of one spin state, reportedly producing a n-spin flip through two processes, both limited by e-spin relaxation. Reliance on e-spin relaxation is a general feature of many nuclear polarization processes, which may be a limiting step in systems with long e-spin relaxation times.

\begin{figure}
\includegraphics{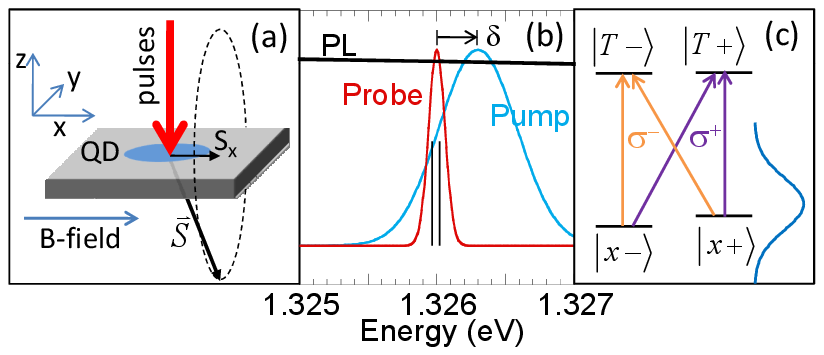}
\caption{\label{fig1} (color online) (a) Experimental geometry showing spin precession. (b) Illustration of pump and probe spectra with the ensemble photoluminescence of the quantum dots. The vertical lines centered on the probe spectrum represent the Zeeman split transitions of a single QD. (c) Electron-trion level diagram, showing the two electron and trion spin states and the allowed transitions. Hole spin splitting is assumed to be negligible, allowing the trion spin states to be written in the optical ($z$) basis.}
\end{figure}

Here, we show that pulses with an intermediate bandwidth can take advantage of both the coherent nature of pulses and the spectral selectivity of narrow-bandwidth lasers. Using two-color time-resolved Faraday ellipticity\cite{Baumberg_PRB94} (TRFE) in an ensemble of InAs QDs, we measure the effect of detuned pulse trains on the e-spin precession frequency, which reflects the degree of nuclear polarization. With only a small asymmetry in transition probabilities from the two spin eigenstates, a significant nuclear polarization of a few percent is achieved, without requiring e-spin relaxation. Instead, optical spin rotation of the e-spin polarization resets the system. By varying the pulse bandwidth, this technique could be used to controllably vary the nuclear polarization and then fix it in a narrow distribution.

\section{Experiment}

This study is performed on an ensemble of $20$ layers of self-assembled InAs QDs grown by molecular beam epitaxy, as described in Ref.~\onlinecite{Carter_PRL09}. Only a narrow spectral distribution of QDs is measured with an optical probe pulse fixed at $1.326$ eV, having a spectral width of 140 $\mu$eV (13 ps). A circularly polarized pump pulse of spectral width $0.7$ meV ($\sim$2 ps) has a variable detuning from the probe (see Fig.~\ref{fig1}(b)). Due to the selection rules for the electron-trion system (Fig.~\ref{fig1}(c)), circularly polarized light generates a superposition of the eigenstates, $\left|x+\right\rangle$ and $\left|x-\right\rangle$, that is nominally oriented along the optical axis. As illustrated in Fig.~\ref{fig1}(a), the e-spin precesses about the perpendicular applied magnetic field (Voigt geometry). The spin polarization is measured with the linearly-polarized probe through TRFE, which measures the difference in absorption between the two circular polarizations as a function of probe delay.



\begin{figure}
\includegraphics{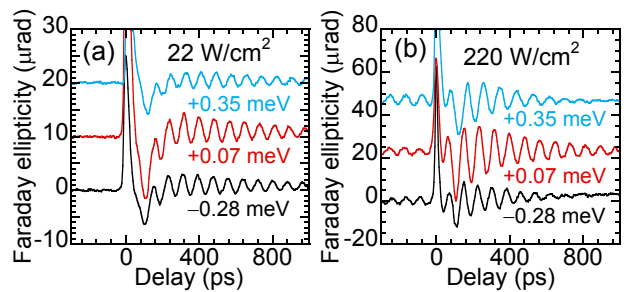}
\caption{\label{fig2} (color online) (a,b) Time-resolved Faraday ellipticity (TRFE) for three pump detunings at (a) low pump intensity of 22 W$/$cm$^2$ and (b) high pump intensity of 220 W$/$cm$^2$. The magnetic field is 2 T. }
\end{figure}

Figure \ref{fig2}(a) displays TRFE at low pump intensity for three pump detunings, $\delta = \hbar\omega_{pump} - \hbar\omega_{probe}$. There are three contributions to the signal: neutral excitons in uncharged dots, trions in negatively charged dots, and electrons in negatively charged dots. We focus on the e-spin signal, which gives the damped oscillations, here at 12 GHz, as they precess in the transverse magnetic field. The $T_2^*$ dephasing time of $\sim$500 ps is primarily due to inhomogeneity in the g-factor of the ensemble, with some contribution from fluctuations in the nuclear polarization. Detuning has little effect at this pump intensity except to decrease the oscillation amplitude.

At a higher pump intensity of $\sim$220 W$/$cm$^2$ in Fig.~\ref{fig2}(b), the oscillation amplitude increases, and a weak signal is observed at negative delays. This signal is due to spin ``mode-locking" as observed previously, \cite{Greilich_Science06, Greilich_Science07, Carter_PRL09, Greilich_PRB09} in which e-spins synchronized to the pulse train are efficiently polarized. A comb of enhanced precession modes at the phase synchronization condition (PSC), $\omega_{PSC} = 2n\pi/T_R$, where $T_R$ is the pulse repetition period, leads to rephasing before each pulse. Spin mode-locking requires a coherence time $T_2$ much longer than $T_R$, which is $12.3$ ns here. Previous results on this sample gave a $T_2$ of 100-200 ns at 3 T and a strong negative delay signal. \cite{Carter_PRL09} The negative delay signal in Fig.~\ref{fig2}(b) is weaker due to a decrease in $T_2$ with decreasing magnetic field. 

\begin{figure}
\includegraphics{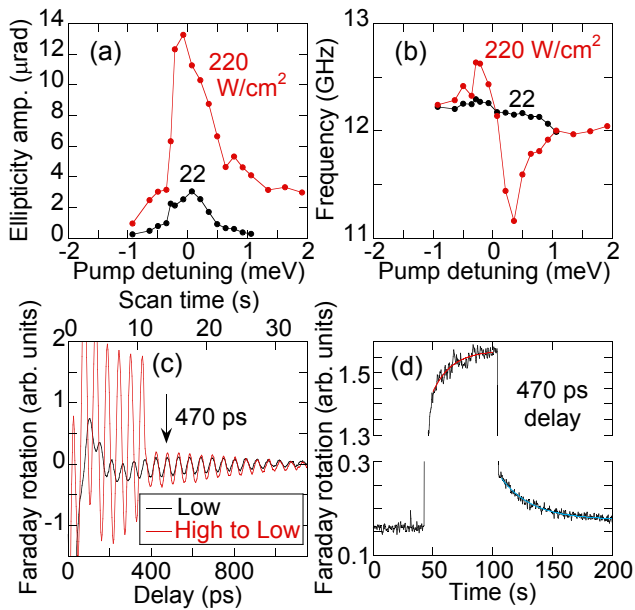}
\caption{\label{fig2p5} (color online) a) Amplitude and (b) frequency of the oscillations obtained from TRFE as a function of pump detuning for low and high pump intensities. (c,d) Time-resolved Faraday rotation (TRFR) at 3 T for degenerate pump/probe with spectrally broad ($0.7$ meV) pump \textit{and} probe pulses. In (c) the pump intensity is reduced from high (600 W$/$cm$^2$) to low (60 W$/$cm$^2$) during the delay scan, with comparison to data taken at low intensity throughout the scan. In (d) the delay is fixed at 470 ps, and the pump intensity is taken from low to high to low. }
\end{figure}

Evidence of nuclear dynamics in this data can be found by examining the amplitude and precession frequency of these oscillations as a function of pump detuning, as plotted in Figs~\ref{fig2p5}(a,b). At high pump intensity there is an asymmetry in the amplitude about zero detuning that was observed and explained in Ref.~\onlinecite{Carter_PRL09}. For positive detuning, nuclear focusing pushes e-spins toward PSCs, giving a strong spin polarization. While for negative detuning, nuclear focusing pushes spins away from PSCs, giving a weaker spin polarization. In Fig.~\ref{fig2p5}(c,d) this asymmetry, as measured by time-resolved Faraday rotation (TRFR), is shown to persist for tens of seconds after reducing the pump intensity.\cite{Carter_SPIE2010} The TRFR signal amplitude is strongly dependent on asymmetry between positive and negative detuning.\cite{Carter_PRL09} When the pump intensity is suddenly decreased during a delay scan (Fig.~\ref{fig2p5}(c)), the oscillation amplitude does not immediately decrease to the amplitude observed for low pump intensity. The fast reduction in amplitude is due to the fast decay of the e-spin polarization, and the slower decay is due to a decay of nuclear focusing. The time constants for buildup (16 s) and decay (26 s) of nuclear focusing are obtained from the data in Fig.~\ref{fig2p5}(d), and these slow dynamics provide strong evidence of nuclear dynamics. The nuclear focusing that gives rise to this asymmetry is due to a narrowing of the nuclear polarization distribution at or in-between the PSCs but does not indicate a net nuclear polarization.

The key result of this article is that these detuned pulses of modest bandwidth also cause a significant shift in the precession frequency due to a net nuclear polarization. The precession frequency is plotted in Fig.~\ref{fig2p5}(b), showing a decrease in frequency at positive detunings and an increase at negative detunings. The maximum frequency change of 1 GHz corresponds to a nuclear field of 170 mT. The origin of the nuclear polarization is the small asymmetry in pump excitation between the two Zeeman split transitions, as illustrated in Fig.~\ref{fig1}(b). For the 2 ps pulses used here, this asymmetry induces a small e-spin component along $x$ that polarizes the n-spins. 


\begin{figure}
\includegraphics{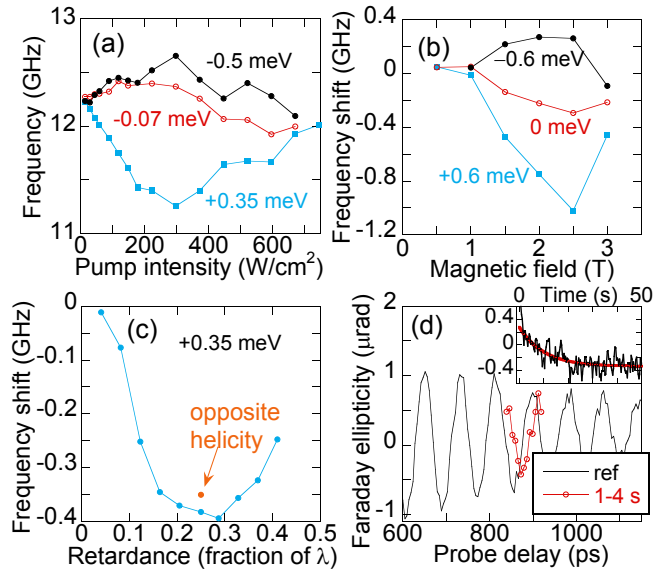}
\caption{\label{fig3} (color online) (a,b,c) Change in e-spin precession frequency as a function of (a) pump intensity at 2 T, circular pump polarization; (b) magnetic field at 600 W$/$cm$^2$, circular polarization; and (c) pump polarization at 3 T, 220 W$/$cm$^2$. A retardance of $\lambda/4$ corresponds to circular, and 0 or $\lambda/2$ correspond to linear. (d) TRFE taken 1-4 seconds after suddenly reducing the pump intensity from 300 to 30 W$/$cm$^2$. For the ``ref" data the pump intensity had been at 30 W$/$cm$^2$ for many minutes. The inset shows the slow change in signal at 850 ps delay due to decay of the nuclear polarization.}
\end{figure}

The change in precession frequency can be controlled with pulse intensity, as displayed in Fig.~\ref{fig3}(a). The shift grows larger with intensity until $\sim$300 W$/$cm$^2$, at which point it starts to decrease. The maximum shift is roughly correlated with saturation of the trion transition. We also expect a strong dependence on magnetic field since the spectral selectivity between spin states will be weaker at smaller Zeeman splittings. In Fig.~\ref{fig3}(b), the frequency shift is essentially zero below 1 T, grows with field until $2.5$ T, and then decreases at 3 T.  The decrease in frequency shift at 3 T is not expected, but a possible explanation will be discussed later. The abrupt decrease to zero shift below 1 T is attributed to the lack of spin mode-locking at this field, perhaps indicating the need for spin coherence lasting in-between pulses.

If the mechanism for this nuclear polarization only relies on exciting a trion from one e-spin state more than the other, one might expect an insensitivity to pump polarization. Figure~\ref{fig3}(c) plots the precession frequency shift as a function of the retardance of the Soleil-Babinet compensator used to change the pump polarization. The shift is maximum for circular polarization ($\lambda/4$ retardance) and approaches zero for linear polarizations (0 and $\lambda/2$). This result is a clear sign that spin coherence and spin rotations, which do not occur for linearly polarized short pulses,\cite{Baumberg_PRB94, Kim_PRL10} are important for developing the nuclear polarization. The helicity of the circular polarization was not observed to have an effect since it should not change the sign of $S_x$.

In Fig.~\ref{fig3}(d), the frequency shift is shown to persist on a time scale of $\sim$10 seconds after suddenly decreasing the pump intensity from 300 to 30 W$/$cm$^2$. This persistence is another clear sign of nuclear effects, and the decay, shown in the inset, represents the loss of nuclear polarization. 

\section{Analysis}

These results can be modeled and understood using a similar treatment as in Ref.~\onlinecite{Carter_PRL09}. Transitions from the e-spin states to the trion states and subsequent recombination are treated as an effective e-spin dephasing $\Gamma_{opt}$ that allows n-spin flips through the electron-nuclear hyperfine interaction.\cite{Greilich_Science07} In Ref.~\onlinecite{Carter_PRL09} the probability to flip up or down $w_{\pm}$ was proportional to the e-spin populations, $\rho_{\pm\pm} = \left(1 \pm 2 S_x \right)$, \cite{DyakonovPerel74} without regard to asymmetry in the excitation probability difference between $\left|x+\right\rangle$ and $\left|x-\right\rangle$ to a trion state. Here, we take this excitation asymmetry into account with $\alpha_{\pm}$, which are the relative excitation probabilities to the trion from $\left|x\pm\right\rangle$. The n-spin flip rates are then 
\begin{equation}
w_{\pm} = \left[A/ \left(\hbar \omega_e N\right) \right]^2 \alpha_{\pm} \Gamma_{opt} \rho_{\pm\pm},
\label{fliprates}
\end{equation}
where $A\approx 100$ $\mu$eV is the hyperfine constant, $N\approx 2\times10^4$ is the number of nuclei, and $\omega_e$ is the e-spin precession frequency. The effective e-spin dephasing $\Gamma_{opt}$ is the rate that trions are excited, $\rho_{TT}/T_R$, where $\rho_{TT}$ is the trion population generated by a pulse. (The trion state is $\left|T+\right\rangle$ or $\left|T-\right\rangle$, depending on the helicity of pulses.) For very short pulses with $\alpha_{\pm} = 1$ Eq.~\ref{fliprates} reduces to that of Ref.~\onlinecite{Carter_PRL09}. The combination of $\alpha_{\pm}$ and the e-spin populations $\rho_{\pm\pm}$ give rise to an asymmetry in nuclear spin flip rates that polarize the nuclei.

Physically we can motivate the $\alpha_{\pm}$ factors by considering the case where a pulse would excite only one of the two transitions, say $\left|x+\right\rangle \rightarrow \left|T\right\rangle$. Then, by performing a Schrieffer-Wolff transformation to the total Hamiltonian with respect to the flip-flop terms, as in Ref.~\onlinecite{Latta_NPhys09}, we get processes such as the following:
\begin{equation}
\left|x+,\Downarrow\right> \stackrel{\mathrm{pulse}}{\xrightarrow{\hspace*{0.8cm}}} \left|T,\Downarrow\right> \stackrel{\left(A/\omega_e\right) \times \mathrm{emission}}{\xrightarrow{\hspace*{2cm}}} \left|x-,\Uparrow\right> ,
\label{flipflop}
\end{equation}
where $\Uparrow$ and $\Downarrow$ are the nuclear spin states. This process, which corresponds to hyperfine-assisted spin-flip Raman scattering, occurs proportional to $\left|U_{Tx+}\right|^2 \rho_{++}$, where $U$ is the pulse evolution operator of the electron-trion system. For a pulse exciting only the $\left|x-\right\rangle \rightarrow \left|T\right\rangle$ transition, a similar process flips $\left|x-,\Uparrow\right>$ to $\left|x+,\Downarrow\right>$ at a rate proportional to $\left|U_{Tx-}\right|^2 \rho_{--}$. Our situation is more complex since the pulse acts on both transitions and does so coherently. To capture the essential physics, we simply weigh the nuclear spin flip rates according to $\alpha_{\pm} = 2\left|U_{Tx\pm}\right|^2 / ( \left|U_{Tx+}\right|^2 + \left|U_{Tx-}\right|^2 )$. This factor takes into account the partial spin selectivity and reduces to $\alpha_+ = \alpha_- = 1$ for short pulses with $U_{Tx+} = U_{Tx-}$. Without this factor, the calculated precession frequency shift is opposite in sign to that observed experimentally. A microscopic study of the exact mechanism of nuclear spin flips is beyond the scope of this paper and will be ideally carried out in conjunction with a single dot experiment, so that crucial information is not washed out by ensemble effects.

\begin{figure}
\includegraphics{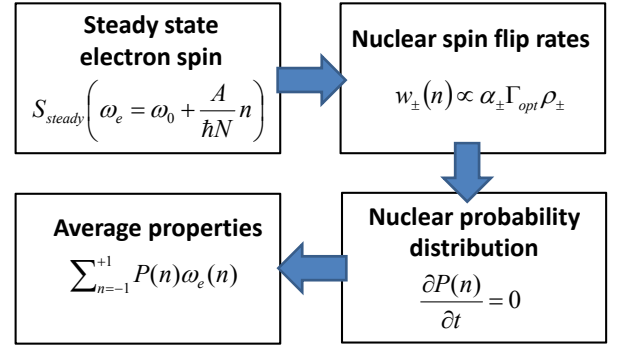}
\caption{\label{fig4p5} (color online) Steps for calculating the nuclear spin polarization and frequency shift. }
\end{figure}

Since the nuclear spin flip rates depend on the e-spin polarization $\bm{S}$, and $\bm{S}$ depends on the nuclear polarization, the electron-nuclear dynamics must be calculated self-consistently. The e-spin dynamics are much faster than changes in the n-spin polarization, so we assume that the e-spin can come to a quasi-steady state while the nuclear polarization is essentially static. The e-spin and n-spin dynamics can then be calculated separately.\cite{Merkulov_PRB02} The steps in calculations are shown in Fig.~\ref{fig4p5}. First, the steady state $\bm{S}$ due to the pulse train is calculated as a function of precession frequency, which corresponds to the possible n-spin polarizations $n$. Second, the n-spin flip rates $w_{\pm}$ as a function of $n$ are determined based on $\bm{S}$ and the pulse properties. Third, the steady state n-spin polarization probability distribution $P\left(n\right)$ is calculated using $w_{\pm}\left(n \right)$. Fourth, properties such as the average nuclear polarization and precession frequency shift are calculated. For simplicity, the calculations are performed for a single QD instead of an ensemble. 


\begin{figure}
\includegraphics{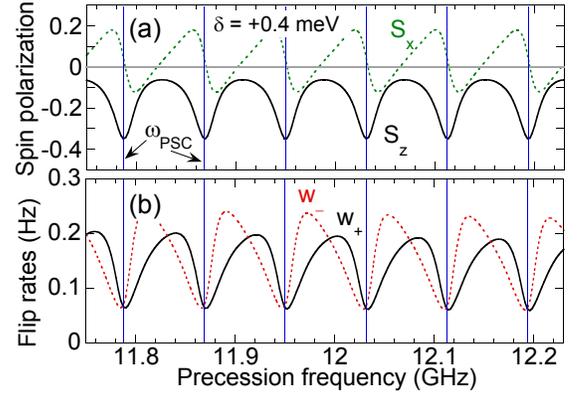}
\caption{\label{fig4} (color online) Calculated (a) steady-state spin polarization and (b) n-spin flip rates as a function of precession frequency for $\pi$-pulses at $\delta=+0.4$ meV. }
\end{figure}


The steady state $\bm{S}$ is determined by first numerically calculating the time evolution operator $U$ of the electron-trion system due to a single hyperbolic secant pulse of bandwidth $0.7$ meV.  This is calculated for a range of precession frequencies and detunings. No decoherence is included during the pulse, and recombination is assumed to return the trion population incoherently.\cite{Shabaev_PRB03, Economou_PRB05, Dutt_PRL05} The evolution operator in-between pulses, which includes an e-spin decoherence time of 100 ns, and the pulse evolution operator yield the steady state $\bm{S}$. An example of $\bm{S}$ just after a pulse is plotted in Fig.~\ref{fig4}(a) for a $\pi$-pulse at $\delta=+0.4$ meV. The result is very similar to that calculated in Ref.~\onlinecite{Carter_PRL09} in the short pulse limit (negligible spin splitting compared to the pulse bandwidth). $S_z$ is enhanced at the PSCs, and $S_x$ alternates between positive and negative, passing sharply through zero near each PSC. With short pulses, $S_x$ is entirely due to optical rotation of $S_y$ about the optical axis, which only occurs away from the PSC. The main difference with a finite spin splitting is that the average $S_x$ (averaging over precession frequency from one PSC to another) is nonzero and depends on detuning. This average $S_x$ is a result of the partial spectral selectivity.


The n-spin flip rates are plotted in Fig.~\ref{fig4}(b) for a positive detuning, showing low flip rates near the PSCs. As explained in Ref.~\onlinecite{Carter_PRL09}, surrounding each PSC, there is a larger $w_+$ below each PSC and a larger $w_-$ above each PSC, leading to focusing toward the PSC. The point of interest here is that the average $w_-$ is greater than the average $w_+$, giving rise to a nuclear polarization. This asymmetry occurs because $\left|x-\right\rangle$ is excited to the trion a bit more than $\left|x+\right\rangle$ ($\alpha_- > \alpha_+$). One might expect that the steady state populations of the spin states would adjust to the different excitation rates to give $\alpha_+ \rho_{++} = \alpha_- \rho_{--}$, eliminating any asymmetry. This is the case for incoherent pumping.  For example, a cw laser resonant with only $\left|x-\right\rangle$ ($\alpha_+ =0$) pumps all the population into $\left|x+\right\rangle$, giving $\alpha_+ \rho_{++} = \alpha_- \rho_{--} =0$. 
Then, no further optical transitions (and nuclear spin flips) can occur without e-spin relaxation, either through the hyperfine interaction or through other mechanisms, which may limit the rate of nuclear polarization. Here, the pulses produce coherent spin rotations about the optical axis, allowing asymmetry in the n-spin flip rates without e-spin relaxation. When rotations and coherence are eliminated in the simulations using linearly polarized pulses, the n-spin flip rates are equal. This explains the reduction in frequency shift in Fig.~\ref{fig3}(c) when approaching linear polarization.


\begin{figure}
\includegraphics{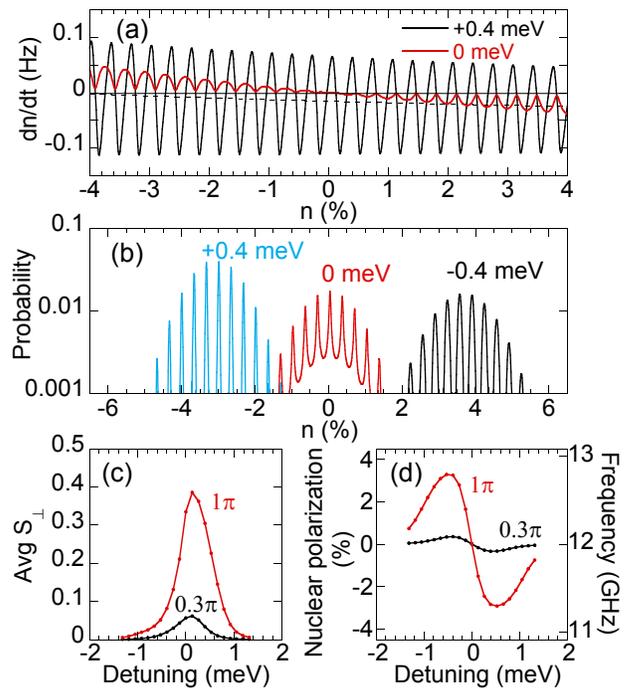}
\caption{\label{fig5} (color online) a) Calculated $d\bar n/dt$ vs. $\bar n$ for $\pi$-pulses at zero and $+0.4$ meV detuning. The dashed line is the detuned data smoothed over the oscillations. (b) Calculated $P(n)$ vs. $n$ for $\pi$-pulses at zero and $\pm 0.4$ meV detuning. (c,d) Calculated average precessing spin (c) amplitude and (d) frequency as a function of pulse detuning for $0.3\pi$ and $\pi$ pulses. }
\end{figure}

The nuclear probability distribution, $P(n)$, is calculated using $w_{\pm}$, which is a function of the nuclear polarization $n = \left( N_{\uparrow} - N_{\downarrow} \right)/N$. It is instructive to first analyze the dynamics of the average nuclear polarization $\bar n$, ignoring the distribution, according to $d\bar n/dt = w_+ - w_- - \bar n \left(w_+ + w_-\right)$. This rate is plotted in Fig.~\ref{fig5}(a) for a positive detuning and zero detuning. For positive detuning, there are stable fixed points near each PSC, where $d\bar n/dt=0$ with a negative slope. For zero detuning, $d\bar n/dt$ approaches zero at the PSCs but never passes through zero, such that the only stable fixed point is at $\bar n =0$. For the detuned case there are many stable fixed points, but the average $d\bar n/dt$, smoothing over the oscillations, is slightly negative (see dashed line). Fluctuations that take $n$ away from a PSC tend to move the system toward negative $n$ until the average rate is zero. 

This effect is observed by calculating the full probability distribution which includes the effects of fluctuations. For simplicity we consider each nucleus to be a spin-$1/2$ system, which captures the essential physics. $P(n)$ evolves according to the following set of $N$ coupled equations
\begin{eqnarray}
\frac{\partial P(n)}{\partial t} &= -N_{\downarrow}P(n)w_+(n) - N_{\uparrow}P(n)w_-(n) \\ 
					 &+ (N_{\uparrow} + 1)P(n^+)w_-(n^+) \nonumber\\
					 &+ (N_{\downarrow} + 1)P(n^-)w_+(n^-), \nonumber
\label{Prob_eqn}
\end{eqnarray}
where $n^{\pm}= n \pm 2/N$ is the polarization higher or lower than $n$ by one flip. Setting Eq.~3 equal to zero gives the steady state nuclear distribution, which is plotted in Fig.~\ref{fig5}(b) for positive, negative, and zero detuning. For positive (negative) detuning, $P(n)$ is centered at (in-between) the PSCs and shifted to a negative (positive) $n$. For zero detuning, $P(n)$ is primarily at the PSCs and centered at $n=0$. The average precessing spin amplitude and frequency resulting from these distributions at a series of detunings are plotted in Fig.~\ref{fig5}(c,d). A small n-spin polarization decay rate ($2\times 10^{-2}$ Hz) was included to prevent the buildup of nuclear polarization at large detunings where the optical excitation rate is negligibly small. This decay rate was chosen to reproduce the observed decay of nuclear polarization at large detunings. The calculations agree reasonably well with the experimental results in Fig.~\ref{fig2p5}(a,b). There is some asymmetry in the measured frequency shifts between positive and negative detuning, which does not agree with calculations. Part of this asymmetry appears to be due to a small decrease in precession frequency at high pulse intensity that is independent of detuning, perhaps due to heating. Another point to consider is that the nuclear polarization requires fluctuations away from the fixed points, which may take a long time. This time dependence is ignored by calculating the steady state $P(n)$ and may be different for positive and negative detuning. These slow dynamics are likely the reason for the decrease in the frequency shift at 3 T (Fig.~\ref{fig3}(b)). At this magnetic field, $T_2$ is longer, which makes the fixed points more stable and fluctuations from point to point less likely.

\section{Conclusion}

The main conclusion of this study is that the bandwidth of optical pulses can be used to obtain more flexible control of nuclear dynamics. Very short pulses have no spectral selectivity but can focus QD spins toward or away from synchronization, stabilizing and narrowing the distribution. Longer pulses with even a little spectral selectivity can shift the n-spin distribution and modify the e-spin precession frequency. These effects do not require e-spin relaxation and instead rely on coherent spin rotations. With improved spectral selectivity (narrower bandwidth pulses) and an improved understanding of the n-spin flip processes, much larger polarizations should be possible. By dynamically controlling pulse bandwidth, it should be possible to shift the nuclear polarization toward a desired value and then fix it in place to give a stable, narrowed distribution. 

This work is supported in part by the US Office of Naval Research. S. E. E. acknowledges support from NSA/LPS, and A. S. acknowledges support from NIST 70NANB7H6138 Am 001.


\bibliography{references}
\end{document}